\documentclass[preprint,12pt]{elsarticle}

\usepackage{amssymb}

\journal{Nuclear Physics B}

\begin{document}

\begin{frontmatter}



\title{Generalized belief propagation for the magnetization of the simple cubic Ising model}


\author{Alessandro Pelizzola}

\address{
Dipartimento di Scienza Applicata e Tecnologia, CNISM and
  Center for Computational Studies, Politecnico di Torino,
Corso Duca degli Abruzzi 24, I--10129 Torino, Italy}
\address{INFN, Sezione di Torino, via Pietro Giuria 1, I-10125 Torino,
Italy}
\address{Human Genetics Foundation, HuGeF, Via Nizza 52, I-10126 Torino,
Italy
}

\begin{abstract}
A new approximation of the cluster variational method is introduced
for the three--dimensional Ising model on the simple cubic lattice. 
The maximal cluster is, as far as we know, the largest ever used in
this method. A message--passing algorithm, generalized belief
propagation, is used to minimize the variational free
energy. Convergence properties and performance of the algorithm are
investigated. 

The approximation is used to compute the spontaneous magnetization,
which is then compared to previous results. Using the present results
as the last step in a sequence of three cluster variational
approximations, an extrapolation is obtained which captures the
leading critical behaviour with a good accuracy. 
\end{abstract}


\end{frontmatter}


\section{Introduction}
\label{sec:intro}

A major approximate tool in the equilibrium statistical physics of
lattice models is the mean--field theory, together with its many
generalizations. These techniques are known to give quite often
reliable qualitative results, which makes them very useful in
understanding properties of a model like its phase diagram. Due to the
limited quantitative accuracy of simple mean--field theory, many
generalizations were developed. Since mean--field neglects
correlations, typically the idea is to include local, but
progressively longer range correlations in the treatment, for example
by means of cluster expansions where clusters of increasing size can be
included. 

This line of research started with the Bethe--Peierls approximation
\cite{Bethe35,Peierls36}, where nearest--neighbour correlations are
taken into account, and the Kramers--Wannier approximation
\cite{KraWan1,KraWan2}, including correlations up to a square
plaquette. Many generalizations were then proposed, and a particularly
successful one was the cluster variational method (CVM), introduced by
Kikuchi in 1951 \cite{Kikuchi51} and applied to the Ising model. The
largest clusters considered by Kikuchi in this work were a cube of 8
sites for the simple cubic lattice and a tetrahedron of 4 sites for
the face centered cubic lattice. 

Larger clusters were later considered: in 1967 Kikuchi and Brush
\cite{KikBru} introduced the $B_{2L}$ sequence of approximations for
the two--dimensional square lattice, whose convergence properties were
later studied \cite{Review} using maximal clusters up to 13
sites. Kikuchi and Brush also suggested that a similar approach could
in principle be carried out in three dimensions, although the
computational costs would have been prohibitively large at that time. 
Their intuition was put on a firmer ground by
Schlijper \cite{Sch83,Sch84,Sch85}, who showed that, for
translation--invariant models in the thermodynamical limit, there exist
sequences of CVM approximations whose free energy converges to the
exact one. For a $d$--dimensional model, the largest clusters to
consider grow in $d-1$ dimensions only, as in a transfer matrix
approach. In 3 dimensions this idea was used by the present author to
develop a CVM approximation for the Ising model on the simple cubic
lattice based on an 18--site ($3 \times 3 \times
2$) cluster \cite{CVPAM4}. 

The main difficulty encountered in trying to enlarge the basic
clusters is the computational cost, which grows exponentially with the
cluster size. More precisely the problem can be written as the
minimization of a free energy whose number of indipendent variables
increases exponentially with the cluster size. A significant amount of
work was then devoted to develop efficient algorithms. The original
iterative algorithm proposed by Kikuchi \cite{Kik74,Kik76,KiKoKa}, the
so--called natural iteration method, is not particularly efficient,
but in certain cases it is provably convergent \cite{Pretti} to a
(maybe local) minimum of the free energy. Faster, provably convergent
algorithms were developed more recently \cite{Yuille,HAK}.

A very important step in the direction of speeding up algorithms for
the minimization of the CVM free energy has been made in 2001, when it
was shown \cite{Yed01} that Belief Propagation (BP) \cite{Pearl}, a
message--passing algorithm widely used for approximate inference in
probabilistic graphical models, is strictly related to the
Bethe--Peierls approximation. In particular, it was shown \cite{Yed01}
that fixed points of the BP algorithms correspond to stationary points
of the Bethe--Peierls variational free energy. This result was later
extended by showing that stable fixed points of BP are (possibly
local) minima of the Bethe--Peierls variational free energy, though
the converse is not necessarily true. This provides us with the
fastest, but not always convergent, algorithm for the minimization of
the Bethe--Peierls free energy. When convergent, BP outperforms the
other algorithms by orders of magnitude (see \cite{Review} for a
detailed comparison). BP was also extended to an algorithm, named
Generalized Belief Propagation (GBP) \cite{Yed01,Yed05}, whose fixed
points are stationary points of the CVM free energy for any choice of
basic clusters. Like BP, GBP is extremely fast but not always
convergent. 

The purpose of the work described here is twofold: we aim both to test
how GBP performs in minimizing a CVM free energy with a very large (32
sites) basic cluster, and to make one more step in the hierarchy of
CVM approximations for three--dimensional lattice models. Working on
the Ising model on the simple cubic lattice as a paradigmatic example,
we follow Schlijper's ideas and enlarge the basic cluster for our CVM
approximation in 2 dimensions only, thus choosing a $4 \times 4 \times
2$ cluster (as far as we know, the largest cluster ever considered in
CVM). We then use a GBP algorithm to minimize the corresponding free
energy in a wide range of temperatures and discuss the accuracy of the
result, focussing in particular on the spontaneous magnetization, in
comparison with state--of--the--art results.

The paper is organized as follows: in Sec.\ \ref{sec:meth} we describe
our CVM approximation and the GBP algorithm we use to minimize it, in
Sec.\ \ref{sec:res} we analyze the performance of the algorithm and
the accuracy of the results for the magnetization, and conclusions are
drawn in Sec.\ \ref{sec:disc}. 


\section{Methodology}
\label{sec:meth}

The CVM is based on the minimization of an approximate variational
free energy, which is obtained from a truncation of a cumulant
expansion of the entropy \cite{Sch83,Review,Mor57,Mor72,An88}.  A
particular CVM approximation is specified by the set $R$ of clusters
one wants to keep in the expansion: the typical choice involves a set
of maximal clusters and all their subclusters. Introducing the
M\"obius numbers $\{a_\alpha, \alpha \in R\}$, defined by
\begin{equation}
\sum_{\beta \subseteq \alpha \in R} a_\alpha = 1 \qquad
  \forall \beta \in R,
\label{MobiusNumbers}
\end{equation}
the variational free energy takes the form
\begin{equation}
{\cal F}(\{p_\alpha, \alpha \in R\}) = \sum_{\alpha \in R} a_\alpha
{\cal F}_\alpha(p_\alpha).
\label{CVMFree}
\end{equation}
Here $p_\alpha$ is the probability distribution for cluster $\alpha$
and 
\begin{equation}
{\cal F}_\alpha(p_\alpha) = \sum_{{s_\alpha}} \left[
  p_\alpha({s_\alpha)} H_\alpha({s_\alpha}) + 
  T p_\alpha({s_\alpha)} \ln p_\alpha({s_\alpha)} \right],
\label{ClusterFree}
\end{equation}
where $s_\alpha = \{ s_i, i \in \alpha \}$ is the configuration of
cluster $\alpha$, $H_\alpha$ its contribution to the Hamiltonian and,
as customary, $T$ is the absolute temperature (Boltzmann's constant
$k_B$ has been set to 1). If $H$ is the Hamiltonian of the model under
consideration, then the condition 
\begin{equation}
H = \sum_{\alpha \in R} a_\alpha H_\alpha(s_\alpha)
\end{equation}
must be satisfied. 
In the following we shall consider the nearest--neighbour Ising model
in zero field, so our Hamiltonian will be
\begin{equation}
H = - \sum_{\langle i j \rangle} s_i s_j, \qquad s_i = \pm 1
\end{equation}
(the coupling constant $J$ has also been set to 1, hence the
temperature will be expressed in units of $J/k_B$).  The splitting of
$H$ into contributions $H_\alpha$ from the various clusters appearing
in the expansion is not unique, several (equivalent) choices are
possible. In the following we distribute $H$ evenly among the maximal
clusters only, so that no Hamiltonian terms appear in the subcluster
free energies.

Our choice for the largest clusters in $R$ is based on Schlijper's
result \cite{Sch83,Sch84,Sch85} that for a $d$--dimensional model one
can improve accuracy by increasing the maximal clusters in $d-1$
dimensions only. On the simple cubic lattice, the elementary cubic
cell of $2 \times 2 \times 2$ sites has been already considered by
Kikuchi \cite{Kikuchi51} and a $3 \times 3 \times 2$ basic cluster
made of 4 elementary cubic cells has been used in \cite{CVPAM4}. The
next step in this sequence is then a $4 \times 4 \times 2$ basic
cluster (32 sites, 9 elementary cubic cells). As far as we know, this
is the largest maximal cluster ever used in a CVM approximation. The
number of possible Ising configurations of such a cluster is $2^{32}
\simeq 4 \cdot 10^9$, which is also the number of independent
variables in our variational problem. More precisely, exploiting
lattice symmetries, this number can be reduced by a factor close to
16, leaving us with $\simeq 2^{28} \simeq 2.5 \cdot 10^8$ independent
variables. In order to deal with such a large number of variables, we
shall use the parent--to--child GBP algorithm \cite{Yed05} to
find stationary points of the variational free energy. This will also
provide a test of convergence and performance of the algorithm in a
very large scale problem.

The parent--to--child GBP algorithm is a message--passing algorithm,
based on iterative equations written in terms of quantities called
messages, which are exchanged between clusters. We shall use the
notation $m_{\alpha \to \beta}(s_\beta)$ for a message going from
cluster $\alpha$ to cluster $\beta$, which is a function of the
configuration $s_\beta$ of the latter. Only clusters $\alpha \in R$
with M\"obius number $a_\alpha \ne 0$ are involved in the
message--passing scheme. Exploiting the lattice translational
invariance in the thermodynamic limit, we shall identify a cluster by
the symbol $l_x l_y l_z$, where $l_x$, $l_y$ and $l_z$ are the lengths
of the cluster in the three spatial directions, in terms of lattice
sites. For instance, a cluster made of a single site will be denoted
by 111, nearest--neighbour pairs in the three directions by 211, 121
and 112 respectively, the elementary cubic cell by 222, and our
maximal cluster by 442. With this notation, it is easy to check that,
according to Eq.\ \ref{MobiusNumbers}, if $R$ includes 442 and all its
subclusters, the only clusters with non--vanishing M\"obius numbers
are the following:
\begin{eqnarray}
a_{442} = 1 & a_{432} = a_{342} = -1 & a_{332} = 1 \nonumber \\
a_{441} = -1 & a_{431} = a_{341} = 1 & a_{331} = -1. 
\end{eqnarray}
Thanks to lattice isotropy, when $l_x \ne l_y$, the clusters $l_x l_y
l_z$ and $l_y l_x l_z$ are equivalent, so we need to consider only six
different clusters, that is six probability distributions, related to
each other by marginalization conditions. In the parent--to--child GBP
algorithm \cite{Yed05}, messages go from a (parent) cluster $\alpha$
to a {\em direct} subcluster (child) $\beta \subset \alpha$, where direct
means that there exist no other cluster $\gamma$ with $a_\gamma \ne 0$
such that $\beta \subset \gamma \subset \alpha$. Hence, in the present
CVM approximation, we will have to introduce only the messages
corresponding to the parent--child pairs listed in Tab.\
\ref{ParentChildren}. 

\begin{table}[h]
\begin{center}
\begin{tabular}{|c|c|}
\hline
Parent & Child \\
\hline\hline
431 & 331 \\
\hline
441 & 431 \\
\hline
432 & 332 \\
\hline
442 & 432 \\
\hline
332 & 331 \\
\hline
432 & 431 \\
\hline
442 & 441 \\
\hline
\end{tabular}
\end{center}
\caption{Parent--child pairs for messages in the parent--to--child GBP
algorithm for our CVM approximation}
\label{ParentChildren}
\end{table}

In the parent--to--child GBP algorithm \cite{Yed05}, cluster
probability distributions at a stationary point of the CVM variational
free energy are written in terms of messages (up to a normalization
constant) as
\begin{equation}
p_\gamma({s_\gamma}) \propto \exp\left[ - H_\gamma({s_\gamma})
  \right] \prod_{\beta \subseteq \gamma} \prod_{\beta \subset \alpha
  \in R}^{\alpha \nsubseteq \gamma} m_{\alpha \to
  \beta}({s_\beta}),
\label{GBP-p-vs-mess}
\end{equation}
where $s_\beta$ denotes the restriction of ${s_\gamma}$ to subcluster
$\beta$. In the above products $\beta$ is any subcluster of $\gamma$
(including $\gamma$ itself) with $a_\beta \ne 0$, while $\alpha$ is
any parent of $\beta$ not contained in $\gamma$. 


Messages are computed by iterating equations derived from the
marginalization conditions which must be fulfilled by the probability
distributions of a parent cluster $\alpha$ and one of its children
clusters $\beta$:
\begin{equation}
p_\beta({s_\beta)} = \sum_{{s_{\alpha \setminus \beta}}}
  p_\alpha({s_\alpha}),
\label{CompConstr}
\end{equation}
where $s_{\alpha \setminus \beta} = \{ s_i, i \in \alpha \setminus
\beta \}$. In the resulting set of equations, messages at (iteration)
time $t$ enter the right--hand side and new messages at time $t+1$ are
obtained on the left--hand side. Writing down the above equations for
all parent--child pairs one realizes that two cases can be
distinguished.

As a first example, consider $\beta = 331$ and $\alpha = 332$ and
write the corresponding probability distributions using Eq.\
\ref{GBP-p-vs-mess}. It can be easily checked that all messages
appearing in the left--hand side, except $m_{\alpha \to
  \beta}(s_\beta)$, will appear also in the right--hand side,
cancelling each other. The resulting equation will then give (up to a
normalization constant) directly $m_{\alpha \to \beta}(s_\beta)$ as a
function of other messages and can be included in an iterative scheme
where messages at step $t$ enter the right--hand side and a new value
for $m_{\alpha \to \beta}(s_\beta)$ at time $t+1$ is obtained.

As a second example, consider now $\beta = 431$ and $\alpha = 432$. In
this case, after cancellations, the left--hand side will contain the
product of $m_{\alpha \to \beta}(s_\beta)$ and two more messages,
specifically those which go from the 332 subclusters of $\alpha$ to
the corresponding 331 subclusters of $\beta$. As a consequence, in
order to evaluate new $432 \to 431$ messages at time $t+1$, one must
have already computed the new $332 \to 331$ messages at time $t+1$
from the corresponding equations. By working out the details for all
messages a partial order emerges among the various computations. At
time $t+1$ one has to compute:
\begin{enumerate}
\item first $431 \to 331$ messages;
\item then $441 \to 431$ and $432 \to 332$ (no definite order between
  them);
\item then $442 \to 432$ messages;
\end{enumerate}
and, independent of the above:
\begin{enumerate}
\item first $332 \to 331$ messages;
\item then $432 \to 431$ messages;
\item then $442 \to 441$ messages.  
\end{enumerate}

We shall close this section with a few technical details about the
implementation of the above scheme. 

In the GBP algorithm, messages are defined up to a normalization
constant. More precisely, for a given $\alpha \to \beta$ parent--child
pair, the messages $m_{\alpha \to \beta}(s_\beta), \forall s_\beta$
can be rescaled by a common constant. As a consequence, in order to
check convergence of the iterative scheme, we need to normalize
messages properly. Several (equivalent) choices are possible, we
normalize them at each iteration by requiring that
\begin{equation}
\sum_{s_\beta} m_{\alpha \to \beta}(s_\beta) = 1
\end{equation}
for all $\alpha \to \beta$ parent--child pairs. 

Iteration proceeds until the condition 
\begin{equation}
\Delta = \sum_{s_\beta} \left[ m_{\alpha \to \beta}^{\rm (new)}(s_\beta) -
    m_{\alpha \to \beta}^{\rm (old)}(s_\beta) \right]^2 < \epsilon
\label{Convergence}
\end{equation}
is met, where {\em new} and {\em old} denote messages at times $t+1$
and $t$ respectively. The actual value of $\epsilon$ will be specified
in the next section, where we discuss the performance of the
algorithm.

Finally, as it often occurs with message--passing algorithms, in order
to achieve convergence it is necessary to damp the iteration. There is
not a unique recipe, and a reasonable tradeoff between convergence and
speed must be looked for in any given problem. Here we found that
convergence was always achieved by replacing, after each iteration,
the new messages with the geometric mean of old and new messages.

\section{Results}
\label{sec:res}

The parent--to--child GBP algorithm described in the previous section
was applied to the simple cubic Ising model in the low--temperature
phase. The inverse temperature $K = T^{-1}$ was varied in the
range 0.223 to 0.436, with a step $\delta K = 0.001$. 

\begin{figure}[h]
\centerline{\includegraphics*[width=0.7\textwidth]{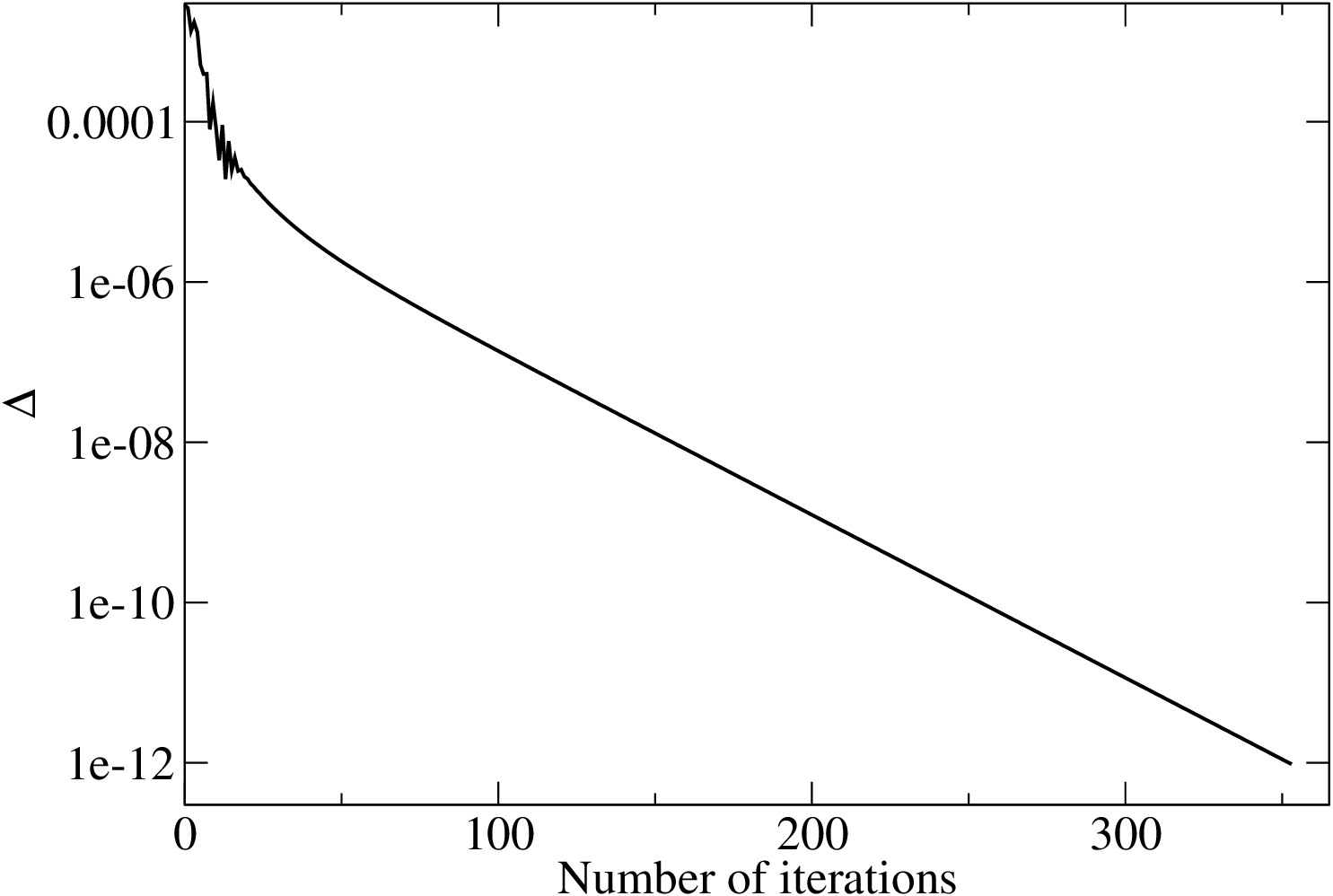}}
\caption{Squared distance $\Delta$ as a function of the number of
  iterations, for $K = 0.223$.}
\label{del_vs_Nit}
\end{figure}

A broken--symmetry initialization was used for the messages:
\begin{equation}
m_{\alpha \to \beta}(s_\beta) = \prod_{i \in \beta} \frac{1 + m_0 s_i}{2},
\end{equation}
with $m_0 = 0.1$. The equations for the messages were then iterated
until the convergence condition Eq.\ \ref{Convergence}, with $\alpha =
442$, $\beta = 441$ and $\epsilon = 10^{-12}$, was met. With this
choice of $\alpha$ and $\beta$ the squared distance $\Delta$ in Eq.\
\ref{Convergence} contains $2^{16}$ terms, so the average rms
variation of individual messages at convergence is not larger than
$10^{-6}/2^8 \simeq 4 \cdot 10^{-9}$. Of course, any other choice of
the $\alpha \to \beta$ parent--child pair (or a combination of all
possible parent--child pairs), with a suitable rescaling of
$\epsilon$, produces similar results. The convergence criterion is
rather strict: as an example, increasing $\epsilon$ to $10^{-9}$ at $K
= 0.24$ affects the spontaneous magnetization in the 7th decimal
place. After a short transient, the squared distance $\Delta$
decreases exponentially with the number of iterations, as illustrated
in Fig.\ \ref{del_vs_Nit}.

\begin{figure}[h]
\centerline{\includegraphics*[width=0.7\textwidth]{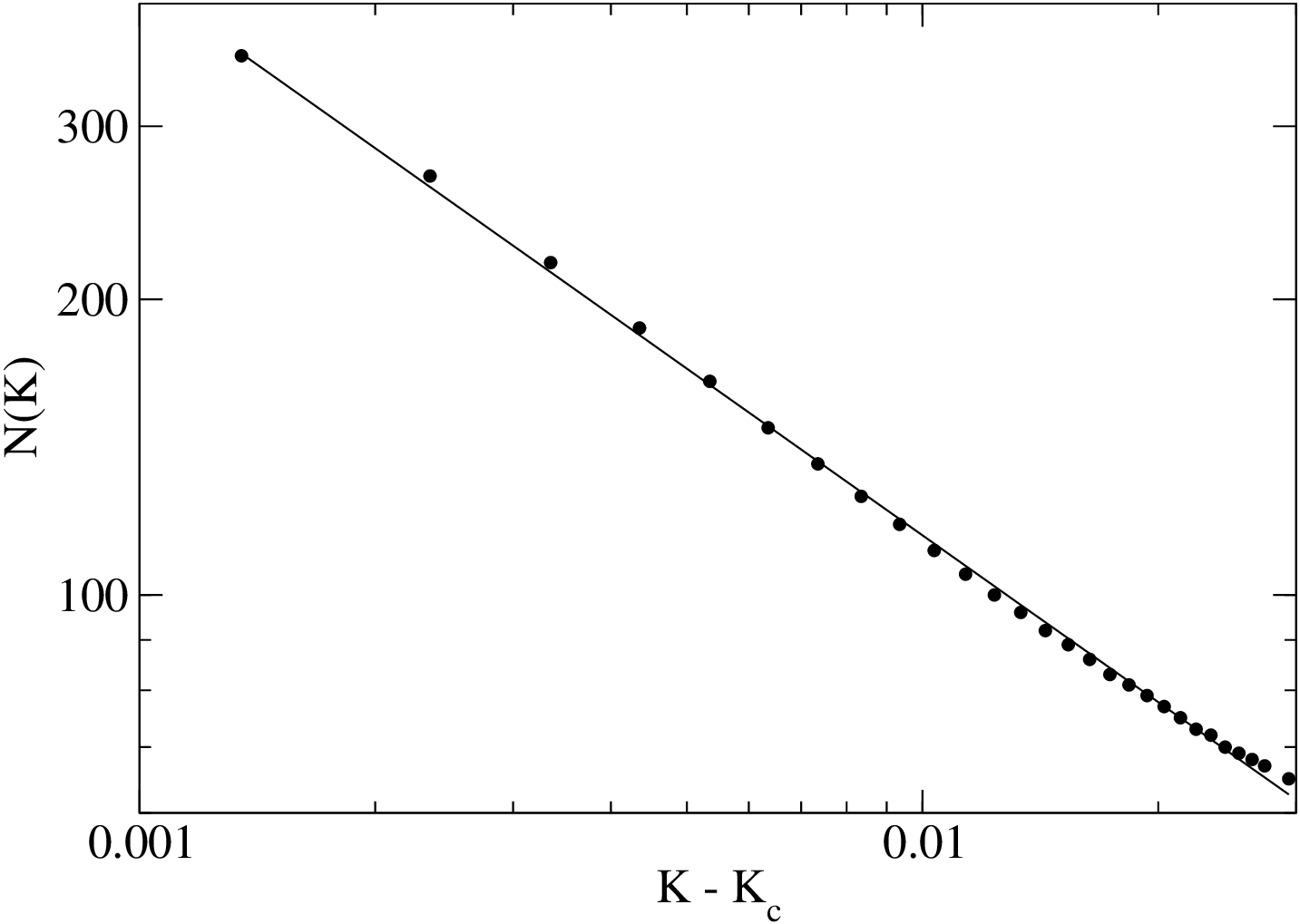}}
\caption{Number of iterations $N(K)$ to reach convergence, as a
  function of $K - K_c$. The solid line is a fit with the
  function $N_0 (K - K_c)^{-z}$.}
\label{SlowingDown}
\end{figure}

The algoritm converges in a number of iterations $N(K)$ which
stays practically constant at 47--48 for $K \ge 0.29$ and exhibits
a critical slowing down as the critical temperature is approached. In
Fig.\ \ref{SlowingDown} we report the number of iterations as a
function of $K - K_c$, where we have used the estimate
$K_c \simeq 0.22165$
\cite{blote,baillie,guttmann,gupta,talapov}. It can be clearly seen
that $N(K)$ is very well fitted by the function $N_0 (K -
K_c)^{-z}$, with $z \simeq 0.564$. In the slowest case, $K =
0.223$, the algorithm took 354 iterations to converge. The number of
messages to be updated at any iteration is mainly determined by the
number of $442 \to 432$ messages, that is $2^{24}$. Exploiting lattice
symmetries this number reduces to $\simeq 2^{22} \sim 4 \cdot
10^6$. For each message, a sum of $2^8$ terms must be evaluated. The
time taken by our code to execute a single iteration on a 2.66 GHz, 64
bits single processor is $\simeq 28$ minutes
, so the
algorithm converges in a time which ranges from approximately 1 day
for $K > 0.29$ to approximately 1 week at $K = 0.223$.

For each $K$, after convergence, we evaluate the probability
distribution of the 331 cluster using Eq.\ \ref{GBP-p-vs-mess}, check
translational invariance, and compute the spontaneous magnetization
$m$. Any correlation function involving a group of sites contained in
the 442 cluster can be computed. 

In order to assess the accuracy of the method, we have compared our
results for the magnetization with the formula by Talapov and Bl\"ote
\cite{talapov}, determined on the basis of high precision simulations
and finite size scaling. For comparison, we have also considered
two lower--order CVM approximation, the cube ($2 \times 2 \times 2$)
one \cite{Kikuchi51} and the 18--site ($3 \times 3 \times 2$) one
\cite{CVPAM4}. In the following we shall denote by $m_{\rm TB}(K)$
the Talapov--Bl\"ote result and by $m_L(K)$ the CVM result from
the approximation with the $L \times L \times 2$ maximal cluster, with
$L$ going from 2 (the cube approximation) to 4 (the present
approximation). 

\begin{figure}[h]
\centerline{\includegraphics*[width=0.7\textwidth]{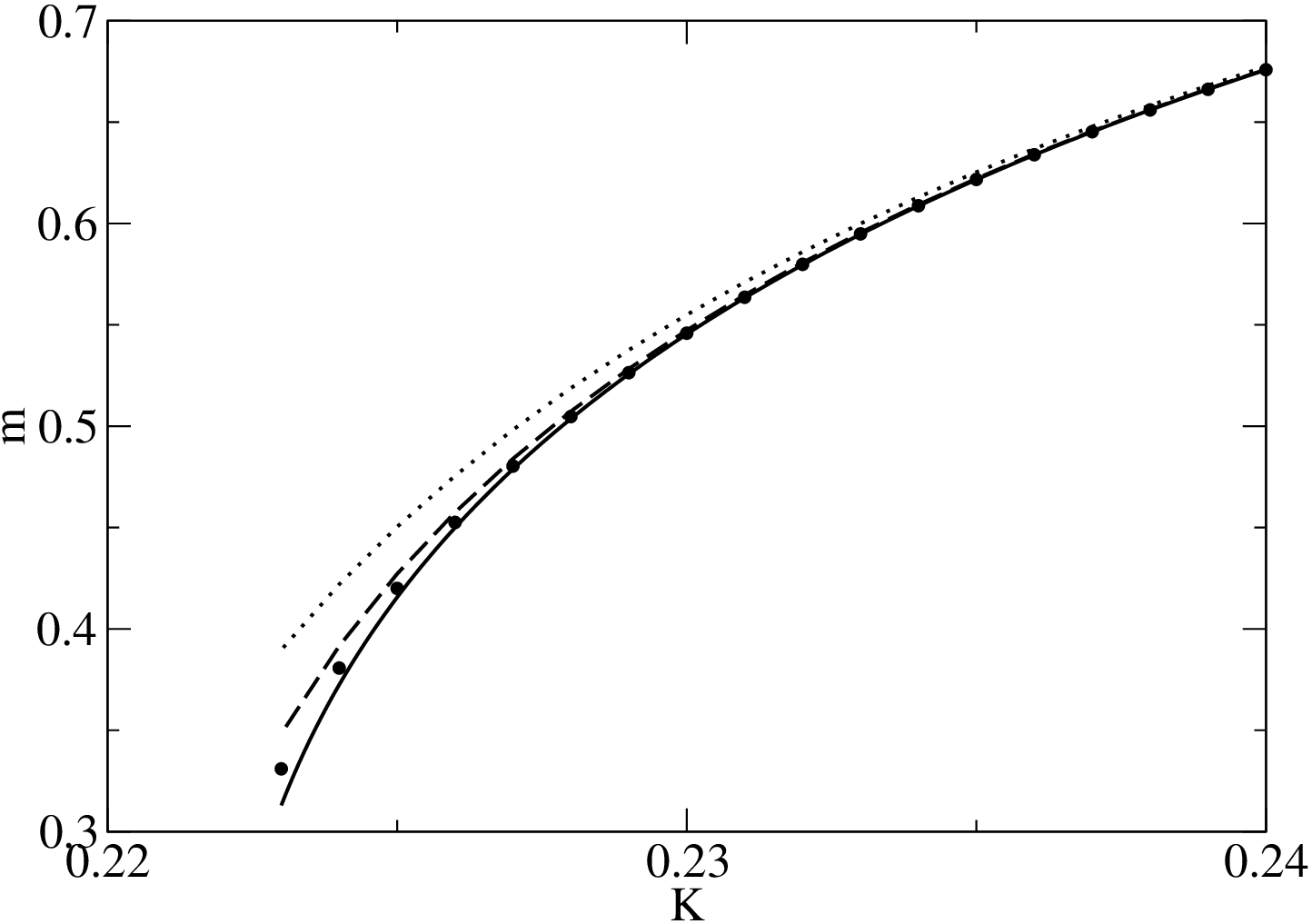}}
\caption{Various estimates of the spontaneous magnetization:
  Talapov--Bl\"ote $m_{TB}$ (solid line) and $m_L$ with $L = 4$
  (present approximation, filled circles), 3 (dashed line) and 2
  (dotted line).}
\label{m234TB}
\end{figure}

A simple plot of the 4 functions in the critical regions is shown in
Fig.\ \ref{m234TB}. The largest deviations occur of course close to
the critical point. In particular, at $K = 0.223$, the present
approximation $m_4$ is larger than the Talapov--Bl\"ote estimate
$m_{TB}$ by $\simeq 0.018$. Corresponding figures for $m_3$ and $m_2$
are 0.035 and 0.077 respectively.

\begin{figure}[h]
\centerline{\includegraphics*[width=0.7\textwidth]{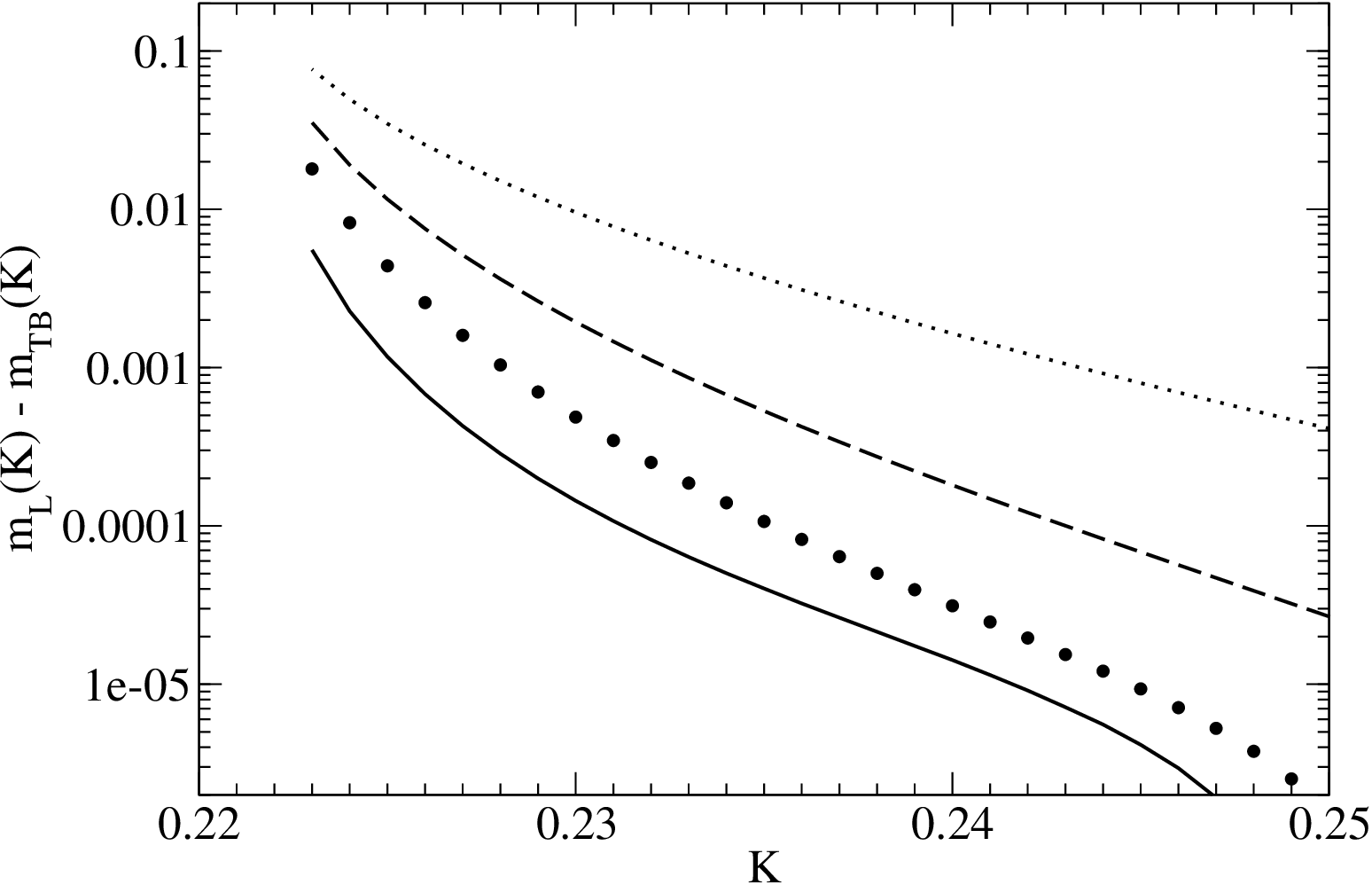}}
\caption{Deviations $m_L(K) - m_{TB}(K)$ from Talapov--Bl\"ote
  estimate, for $L = 4$ (present approximation, filled circles), 3
  (dashed line), 2 (dotted line) and $\infty$ (extrapolation, solid
  line).}
\label{Deviations}
\end{figure}

The above result is better appreciated by plotting the deviations
$m_L(K) - m_{TB}(K)$ from the Talapov--Bl\"ote estimate,
reported in Fig.\ \ref{Deviations}. In this figure we also report the
deviation $m_\infty(K) - m_{TB}(K)$ for an extrapolation
$m_\infty$. In principle, based on Schlijper's results
\cite{Sch83,Sch84,Sch85}, for any $K > K_c$ one would like to
define $m_\infty(K) = \displaystyle{\lim_{L \to \infty}} m_L(K)$, which
should be equal to the exact result. The terms of this sequence for $L
> 4$ are not available, so we need a finite--size ansatz to
extrapolate $m_\infty$ from the results for $L = 2, 3$ and 4. Since
for $K > K_c$ the model has a characteristic length
$\xi(K)$, the correlation length, a natural assumption is that
\begin{equation}
m_L(K) = m_\infty(K) + \delta m(K) \exp\left[ -
  L/\xi(K) \right],
\end{equation}
at least asymptotically. Assuming equality for $L = 2, 3$ and 4 one
finds 
\begin{eqnarray}
m_\infty &=& \frac{m_2 m_4 - m_3^2}{m_2 - 2 m_3 + m_4} \\
\xi &=& \left[ \log(m_2 - m_3) - \log(m_3 - m_4) \right]^{-1}. \label{xi}
\end{eqnarray}
Notice that $m_\infty$ and $\xi$ are independent of an offset which
might be added to (or subtracted from) $L$. For example, one could
measure the size of the maximal clusters in terms of lattice spacings
and obtain $L^\prime = 1, 2$ and 3 for the cube, 18--site and the
present approximation, but this would affect only the value of the
prefactor $\delta m$. It is also important here to stress that without
$m_4$ this extrapolation would not have been possible.

The correlation length $\xi$ from Eq.\ \ref{xi} is strongly affected
by numerical uncertainties, due to the small differences between $m_2,
m_3$ and $m_4$. Indeed, it oscillates, and then becomes not defined,
for $K > 0.26$.  On the other hand, the extrapolated spontaneous
magnetization $m_\infty$ does not suffer from these numerical problems
and it is remarkably accurate. At $K = 0.223$, it is larger than
$m_{TB}$ by 0.05 only, less than $1/3$ the corresponding deviation of
$m_4$. Since $m_\infty(K)$ is our best estimate for the
spontaneous magnetization, it is worth investigating its critical
behaviour. We have fitted our dataset to the function \cite{talapov}
\begin{equation}
m(t) = t^\beta \left( a_0 - a_1 t^\theta - a_2 t \right),
\label{Critical}
\end{equation}
where $t = 1 - K_c/K$ denotes reduced temperature, with 6 fitting
parameters: $K_c, \beta, a_0, a_1, a_2, \theta$ (from now on $\beta$
denotes the critical exponent of the spontaneous magnetization),
obtaining:
\begin{eqnarray}
&& K_c = 0.221510(4), \nonumber \\
&& \beta = 0.332(1), \nonumber \\
&& a_0 = 1.70(1), \nonumber \\
&& a_1 = 0.73(7), \nonumber \\
&& a_2 = 0.03(8), \nonumber \\
&& \theta = 0.72(4). \nonumber
\end{eqnarray}
Given the small value of $a_2$ we also made a similar fit imposing
$a_2 = 0$, with the results
\begin{eqnarray}
&& K_c = 0.221512(2), \nonumber \\
&& \beta = 0.3315(3), \nonumber \\
&& a_0 = 1.694(2), \nonumber \\
&& a_1 = 0.752(2), \nonumber \\
&& \theta = 0.738(2). \nonumber
\end{eqnarray}
For a comparison we recall that Talapov and Bl\"ote best estimate
\cite{talapov}, obtained with $K_c = 0.2216544$, was 
\begin{eqnarray}
&& \beta = 0.3269(3), \nonumber \\
&& a_0 = 1.692(4), \nonumber \\
&& a_1 = 0.344(6), \nonumber \\
&& a_2 = 0.426(11), \nonumber \\
&& \theta = 0.508(15). \nonumber
\end{eqnarray}

We see that the leading term is captured reasonable well by our
approximation, with a slightly smaller $K_c$, a slightly larger
exponent $\beta$ and a compatible prefactor $a_0$, while the same is
not true for the correction to scaling terms. 

\section{Discussion}
\label{sec:disc}

The present paper discusses the application of the CVM approximation
with a $4 \times 4 \times 2$ maximal cluster to the three--dimensional
Ising model on the simple cubic lattice. The maximal cluster is, as
far as we know, the largest ever considered (32 sites) and the
approximation can be viewed as the third step of a sequence of $L
\times L \times 2$ approximations, where $L = 2$ is the original cube
approximation \cite{Kikuchi51} and $L = 3$ was considered in
\cite{CVPAM4}. 

Due to the large size of the maximal cluster, it is necessary to
resort to an instance of the GBP algorithm for the minimization of the
variational free energy. As a consequence, this work also tests the
GBP algorithm with a very large maximal cluster, showing that
convergence can be achieved in reasonable times even with $2^{22}$
messages. The accuracy of the results is assessed by evaluating the
spontaneous magnetization and comparing it with previous
approximations and with a recent best estimate.

In addition to the expected improvement with respect to $L = 2$ and 3,
we observe that the availability of results for three $L$ values
allows an extrapolation, with the only assumption that the approach to
the exact value is exponential in $L$. This assumption is much weaker
than those underlying other techniques used to attempt to extrapolate
non--classical critical behaviour from generalized mean field
theories, like the Cluster Variational -- Pad\'e Approximant Method
\cite{CVPAM1,CVPAM2} or the Coherent Anomaly Method \cite{CAM}. 

The extrapolation gives a good estimate of the leading term of the
critical behaviour, although it cannot reach the accuracy of the
recent best estimates (see e.g.\ \cite{PelissettoVicari} for a review).

It does not seem feasible, at least at the moment, to investigate the
next approximation in the sequence, corresponding to $L = 5$. This
would mean to use a 50--site maximal cluster, making the number of
variables increase by a factor $2^{18}$ with respect to the present
study. 

It would instead be interesting to consider so--called improved Ising
models \cite{Blote3rd,Hasenbusch}, where third--neighbour interactions
are included in order to minimize subleading corrections to scaling.

Finally, it is worth mentioning that while the above analysis was carried out by considering the zero--field magnetization, thus leading to an estimate for the critical exponent $\beta$, it could be extended to several other quantities. The magnetization itself could be computed in the presence of an external field, yielding estimates for the critical isotherm and its exponent $\delta$. Moreover, any correlation function involving a group of
sites contained in our largest cluster can be computed, in particular short--range correlation functions, and hence the internal energy, even in the above--mentioned improved Ising models with third--neighbour interactions. From the magnetization in non--zero field and the internal energy, taking numerical derivatives, response functions like specific heat and susceptibilities can be obtained, and the respective exponents $\alpha$ and $\gamma$ could be estimated.








\end{document}